\documentclass{article}
\def\theequation{\arabic{section}.\arabic{equation}}
\def\r{r^2_0}
\usepackage{bm}
\begin{document}
\def\bbox{\vrule height2mm depth0mm width5pt}
\def\theequation{\arabic{equation}}
\font\sqi=cmssq8
\def\DR{\rm I\kern-1.45pt\rm R}
\def\DC{\kern2pt {\hbox{\sqi I}}\kern-4.2pt\rm C}
 \pagenumbering{arabic}
\setcounter{page}{1}
\newcommand{\beq}{\begin{equation}}
\newcommand{\eeq}{\end{equation}}
\newcommand{\bea}{\begin{eqnarray}}
\newcommand{\eea}{\end{eqnarray}}
\newcommand{\bc}{\begin{center}}
\newcommand{\ec}{\end{center}}
\newcommand{\mb}{\mbox{\ }}
\newcommand{\bs}{\mbox{\boldmath $\sigma$}}
\newcommand{\bp}{\mbox{\boldmath $\pi$}}
\newcommand{\bet}{\mbox{\boldmath $\eta$}}
\newcommand{\bex}{\mbox{\boldmath $\xi$}}
\newcommand{\oQ}{\overline Q}
\newcommand{\bQ}{\mbox{\boldmath $Q$}}
\def\r{r^2_0}
\newcommand{\ch}{{\tt h}}
\newcommand{\ra}{\rightarrow}
\newcommand{\la}{\leftarrow}
\newcommand{\IR}{\mbox{I \hspace{-0.2cm}R}}
\newcommand{\IN}{\mbox{I \hspace{-0.2cm}N}}
\newcommand{\ol}{\overline}
\newcommand{\ul}{\underline}
\begin{center}
{\large \bf Hamiltonian reduction and supersymmetric mechanics with Dirac monopole}\\
\vspace{0.5 cm}
{ \large Stefano Bellucci$^1$,    Armen Nersessian$^{2,3}$ and
Armen Yeranyan$^2$ }
\end{center}
{\it $^1$ INFN-Laboratori Nazionali di Frascati,
 P.O. Box 13, I-00044, Italy\\
$^2$ Yerevan State University, Alex  Manoogian St., 1, Yerevan,
375025, Armenia\\
$^3$ Artsakh State University, Stepanakert \& Yerevan Physics
Institute, Yerevan,
 Armenia}
\date{today}
\begin{abstract}
\noindent We apply the technique of Hamiltonian reduction for the construction of
 three-dimensional ${\cal N}=4$
supersymmetric mechanics specified by the presence of a Dirac
monopole. For this purpose we take the conventional  ${\cal N}=4$
supersymmetric mechanics on the four-dimensional
conformally-flat spaces and perform its Hamiltonian reduction to
three-dimensional system. We formulate the final system in the canonical coordinates,
and present, in these terms,
the explicit expressions of the Hamiltonian and supercharges. We show
that, besides a magnetic monopole field, the resulting system is specified by the presence
 of a spin-orbit
coupling term. A comparison with previous work is also carried out.
\end{abstract}
\noindent
\subsection*{Introduction}
The Hamiltonian reduction appears as an effective procedure for studying the qualitative
properties of classical systems. Also, it is one of the most powerful methods for the
construction of nontrivial integrable systems in classical mechanics.
In fact, all known integrable models of classical mechanics, including multiparticle ones,
could be obtaned by an appropriate Hamiltonian reduction from higher-dimensional trivial integrable
mechanical systems (free-particle and oscillator) \cite{perelomov}.
A specific, particular case of the Hamiltonian reduction is  the reduction of four-dimensional
mechanical systems by the Hamiltonian action of the $U(1)$ group,
which yields the three- dimensional mechanical systems specified by the presence of Dirac monopole.
The best known application of this procedure is the construction of the so-called MIC-Kepler system
(which is the generalization of the three-dimensional Coulomb problem specified by the presence of Dirac
monopole \cite{zwanziger}) from the four-dimensional oscillator \cite{MICred}.
In a similar way the generalization of the MIC-Kepler system
on the three-dimensional hyperboloid  has been constructed
from  the
 oscillator on the four-dimensional sphere and hyperboloid (this system has been suggested in \cite{higgs})
  \cite{np} and
from the oscillator on two-dimensional complex projective space
$\DC P^2$ and  Lobachevsky space
${\cal L}_2$ \cite{cone}. Notice that the appearence in the
reduced system of the Dirac monopole is the result of this
specific reduction procedure, and it has no any direct relation
with the structure of the initial Hamiltonian. Particularly, one
can apply this reduction procedure to the supersymmetric
Hamiltonian systems, and, reducing the number of its bosonic
variables, to obtain the three-dimensional supersymmetric
Hamiltonian system with Dirac monopole. Such an approach to the
construction of three-dimensional  supersymmetric mechanics
looks quite attractive. This is because standard (and powerful)
  approaches to the construction of the  systems with extended supersymmetries are
based on the superfield technique. The latter is related with the complex structures,
and, as a consequence, the configuration spaces of the
corresponding supersymmetric mechanics are K\"ahler or quaternionic spaces.
For example,  the
${\cal N}=4$ supersymmetric mechanics constructed by the use of chiral superfields,
has $(n|2n)_{\DC}$-dimensional configurational superspace,
underlied by the the K\"ahler manifold (see, e.g.\cite{n4bn}), and the
${\cal N}=8$ supersymmetric mechanics constructed by the use of chiral superfields,
has $((n|4n)_{\DC}$-dimensional configurational superspace,
underlied by the special K\"ahler manifold \cite{n8bkn}.
The ${\cal N}=4$ supersymmetric mechanics constructed by the use of the so-called
``root"  supermultiplets \cite{gates} possesses a $(2n.2n)_{\DC} $
-dimensional configuration space,
which is conformally-flat for $n=1$ \cite{root}.
There exists the model of ${\cal N}=8$ supersymmetric mechanics
with $(2n.4n)_{\DC} $-dimensional
configuration space, which is also conformally-flat for $n=1$
\cite{sutulin} (the $n>1$ case has been suggested in \cite{is}).
Also, one can increase the number of supersymmetries, passing from
K\"ahler spaces to hyper-K\"ahler ones, without expanding the number
of fermionic degrees of freedom \cite{korea,krivonos}. Although
supersymmetric mechanics with a Dirac monopole is known
 in the literature (see, e.g. \cite{horvathy,dejohnge,hong} and references therein),
they where found, in some sense, occasionally. While the regular superfield approach
to supersymmetric mechanics does not give the way to incorporate
in the system the interaction with external gauge fields without breaking
supersymmetry.
 Probably, the only exceptions are the three-dimensional ${\cal N}=4$ superconformal mechanics
 \cite{ivanov}
 and the two-dimensional ${\cal N}=4,8$ supersymmetric mechanics constructed within the
``nonlinear chiral superfield" approach \cite{ncs}.
However, it is unclear how to construct nontrivial
higher-dimensional analogs of these systems.
The Hamiltonian reduction could be useful also in this subject, i.e. in the construction of the
even-dimensional  supersymmetric mechanics interacting with gauge fields.

The idea to construct supersymmetric mechanics by the Hamiltonian
reduction is, in some sense, part of the physics folklore.
Explicitly  it was written down, e.g., in \cite{losev}, and
exemplified there by the concrete example of one-dimensional
${\cal N}=4$ supersymmetric mechanics, constructed by the
reduction of the two-dimensional one based on a chiral superfield
(this reduction was performed for the first time in \cite{pashnev}).
Naturally, there is no interaction with non-trivial gauge fields
in this system. Another example is the five-dimensional
supersymmetric mechanics constructed in \cite{smilga}. Let us also
mention the old paper \cite{khudaverdian}, where  the complex
projective superspaces were constructed as reduced phase spaces
of super-Hamiltonian systems. The Hamiltonian reduction seems to
be a natural procedure for the construction of supersymmetric
mechanics including the interaction with external gauge fields
from higher-dimensional supersymmetric systems (without external
gauge fields) constructed within the superfield approach.

In the present paper we demonstrate this fact on the simple case of the reduction of the
$(2|2)_{\DC}$-dimensional  ${\cal N}=4$ supersymmetric mechanics with conformally-flat
configuration space to the three-dimensional system.
In some sense, the content of the presented paper can be considered
the Hamiltonian counterpart of an earlier work \cite{root}.
There it was considered the {\sl Lagrangian} reduction of the four-dimensional
 ${\cal N}=4$ supersymmetric mechanics mechanics constructed
by the use of the ``root" supermultiplet to the three- and two-dimensional  systems.
It was also observed there
that the resulting two-dimensional system coincides with that constructed
by the use of the nonlinear chiral multiplet.
The appearance of the Dirac monopole field
has been detected in the three-dimensional system, and that of the constant magnetic field
was seen in the two-dimensional one.

However, the present paper contains some new features.
 Our resulting  system is formulated
purely in three-dimensional terms and canonical Poisson brackets,
so that passing to supersymmetric
{\sl quantum} mechanics is straightforward here. This formulation allows us to clarify
 the nature of the resulting system. Particularly, we indicate the appearance of
 the spin-orbit coupling term there.
Even when the configuration space of the reduced system is the Euclidean space,
and in the absence of a
 magnetic monopole field, the Hamiltonian of the system contains non-zero fermionic terms.
  Also, in contrast with \cite{root}, we get the
 the supercharges of the reduced system as well, and find that they possess a quite unusual
structure, which seems not to be predictable from current intuition.
Finally,
in our consideration the  odd coordinates of the reduced system
are singular in the coordinate origin only, in contrast with \cite{root},
where  they are singular in the ``Dirac
string", i.e. on a semiaxis. As a consequence, in the previous
consideration, the reduced supercharges and the Poisson brackets
are also singular on the Dirac string, whereas in the present
picture all the ingredients have singularities on the coordinate
origin only.

Before going into details, let us briefly
present our procedure of
Hamiltonian reduction.
Let the initial phase superspace be parameterized
by the local complex coordinates
$(z^{\alpha};\; \pi_{\alpha};\;\eta^{\alpha})$, $\alpha, \beta =1,2$.
 For the Hamiltonian reduction by the action of the constant of motion $J_0$,
  we should find  another set
of coordinates $(x_{i}, \; u;\; p_{i}; \; \xi^{\alpha})$, where
\beq \{J_0,p_{i}\}=\{J_0 ,x_{i}\}=\{J_0,\xi^{\alpha}\}=0,\qquad i,j,k=1,2,3\;. \label{condition} \eeq
The latter coordinate $u$, necessarily has a non-zero Poisson
bracket with ${\cal J}$ (because the Poisson brackets are
non-degenerate )
$\{u,J_0\}\neq 0$.
Then, we
immediately get that in these coordinates the Hamiltonian is
independent of $u$ \beq \{J_0,{\cal H}\}=
\frac{\partial{{\cal H}}}{\partial u}\cdot \{u,J_0\}\neq 0,
\quad\Rightarrow \quad {\cal H}={\cal H}(J_0, x_{i}, p_{i},
\xi^{\alpha}). \eeq
From the Jacobi identity
we get that all Poisson brackets for the phase superspace
coordinates $(x_{i}, \;  p_{i}, \; \xi^{\alpha})$ are also independent on $u$.
 Since $J_0$ is a constant of motion, we can fix
its value $J_0=c$,  and describe the
system in terms of the local coordinates $(x_{i}, \;  p_{i},
\; \xi^{\alpha})$ only. In this way we shall  reduce the initial super-Hamiltonian system with
a $(8|8)_{\DR}$-dimensional phase superspace  to the system with a $(6|8)$-dimensional one.
Geometrically, this  Hamiltonian reduction  means that we fix, in the phase superspace,
the $(7|8)_{\DR}$- dimensional level surface ${\cal M}_{\gamma}$ by the $J_0=c$,
and then  factorize it by the action of a vector field $\{J_0,\;\}$,
 which is tangential  to ${\cal M}_{\gamma}$.
The resulting space ${\cal M}_r={\cal M}_{\gamma}/\{J_0,\;\}$ is a phase superspace of the
 reduced system.

\subsection*{Flat case}

Firstly, we  consider the simplest case of a
${\cal N}=4$ supersymmetric free   particle with {\sl
four} fermionic degrees of freedom, moving on $\DR^4=\DC^2$
equipped with a Euclidean metrics $ ds^2=dz^\alpha d\bar{z}^\alpha $.

This system can be conveniently described in terms of a
$(4|2)_{\DC}$-dimensional phase superspace equipped with the
canonical Poisson bracket. The latter is defined by the following
non-zero relations (and their complex-conjugates): \beq
\{\pi_\alpha , z^\beta \}=\delta^{\beta}_\alpha , \quad
\{\bar{\eta}^\beta,\eta^\alpha\}= \delta^{\bar{\beta}\alpha}.
\label{poisf} \eeq
In order to construct the system with ${\cal
N}=4$ superalgebra
\beq \{Q_\alpha
,\overline{Q}_\beta\}=2\delta_{\alpha\bar{\beta}} {\cal H}, \quad
\{Q_\alpha ,Q_\beta\}=0, \label{algebr} \eeq
we choose the
following supercharges:
 \beq Q_1=\pi_1\eta^1+\bar{\pi}_2\bar{\eta}^2,
\quad Q_2=\pi_2\eta^1-\bar{\pi}_1\bar{\eta}^2,\label{scharg} \eeq
which obey the second equation in (\ref{algebr}). Then,
``squaring" these supercharges (with respect to the Poisson bracket) we
get the ${\cal N}=4$ supersymmetric Hamiltonian \beq {\cal
H}=\bar{\pi}_{\alpha}\pi_{\alpha}/2\;.\label{h} \eeq
Let us notice that the supercharges look quite
simple in the quaternionic notation \beq \bQ
=Q_1-jQ_2={\bp}{\bet},\quad {\bp}=\pi_1-j\pi_2 , \quad
{\bet}=\eta^1+\eta^2 j\;. \eeq
Clearly, the free-particle Hamiltonian  and the supercharges are invariant under  $U(1)$ rotations
\beq
 \quad\delta z^\alpha=\imath z^\alpha\;,\; \delta \eta^\alpha=\imath \eta^\alpha\; ,\;
\delta \pi_\alpha=-\imath\pi_\alpha, \label{sim0}\eeq given by the generator \beq
J_0=\imath(z\pi-\bar{z}\bar{\pi})+\imath \eta\bar{\eta}:\quad
\quad \{J_0 , Q_\alpha \}=\{J_0 , {\overline Q}_\alpha \}= \{J_0 ,
{\cal H}\}=0. \label{generator}\eeq Hence, performing the
Hamiltonian reduction by the action of $J_0$, we shall
get three-dimensional ${\cal N}=4$  supersymmetric mechanics.

Also, we can define the generators of $SU(2)$ rotations acting
separately on bosonic and fermionic variables, which also
commute  with the generator $J_0$
 \bea
&&{\cal J}_{i}=\frac{\imath(z\widehat\sigma_i\pi-\bar{\pi}\widehat\sigma_i\bar{z})}{2}\;:\;
\{J_0,{\cal J}_{i}\}=0,
 \;\{{\cal J}_{i},{\cal J}_{j}\}=-\varepsilon_{ijk}{\cal J}_{k},\label{j}\\
&&\delta_i z^\alpha=\frac{\imath}{2}(z\widehat\sigma_i)^\alpha\;,\;
 \delta_i \pi_\alpha=-\frac{\imath}{2}(\widehat\sigma_i
\pi)_\alpha \;,\;\delta_i \eta^\alpha=0, \eea
 and
  \beq
R_{i}=\frac{\imath}{2} \eta\sigma_i\bar{\eta}\;:\;
\{J_0,R_{i}\}=0, \;\{R_{i},R_{j}\}=-\varepsilon_{ijk}R_{k}, \;
\delta_i z^\alpha=\delta_i \pi_\alpha=0,\; \delta_i
\eta^\alpha=\frac{\imath}{2}(\eta\widehat\sigma_i)^\alpha\;,\label{r}
\eeq
where
$\widehat\sigma_i$ denote Pauli matrices.\\
Notice that these $SU(2)$ generators commute with the Hamiltonian
\beq
\{{\cal J}_{i},{\cal H}\}=0,\quad
\{{R}_{i},{\cal H}\}=0
\eeq
but do not commute with the supercharges

\beq
\{{\cal J}_{i},Q_{\alpha}\}=-\frac{\imath}{2}(\widehat\sigma_i Q)_\alpha\;,\quad
\{R_{1}-\imath R_2,Q_{\alpha}\}=-\imath\varepsilon_{\alpha\beta}\bar Q_\beta,\;
\{R_1+\imath R_2, Q_{\alpha}\}=0,\;
\{R_{3},Q_{\alpha}\}=\frac{\imath}{2}Q_{\alpha}.
\eeq

Hence, performing their
 reduction by the Hamiltonian action of $J_0$, we shall get the
three-dimensional generators of $SU(2)$ rotations of
${\cal N}=4$  supersymmetric mechanics which form, with the supercharges, a  nontrivial superalgebra.\\

Now, let us perform the Hamiltonian reduction.
For this purpose we should fix the $(7|4)_{\DR} $-dimensional level surface of the $J_0$
generator
\beq J_0=2s, \label{const} \eeq
and then factorize it by the $U(1)$-group action given
by the tangent vector field $\{J_0,\;\}$.
The resulting $(6|4)_{\DR}$-dimensional phase superspace
could be parameterized by the following functions:
 \beq p_{i}=\frac{z
\widehat\sigma_{i}\pi+\bar{\pi}\widehat \sigma_{i}\bar{z}}{2 z \bar{z}}, \quad
x_{i}=z \sigma_i\bar{z}, \label{coord}\eeq and \beq
\xi^1=\frac{{z}^1\bar\eta^1+\bar z^2{\eta}^2}{|z|},\quad
\xi^2=\frac{{z}^2\bar\eta^1-\bar z^1{\eta}^2}{|z|}, \label{odd}
\eeq which are clearly $U(1)$-invariant \beq
\{J_0,p_{i}\}=\{J_0, x_{i}\}=\{J_0,\xi^\alpha\}=0. \eeq
The reduced bosonic coordinates (\ref{coord}) are exactly the same,
 which appear in the Kustaanheimo-Stiefel transformation of the four-dimensional
 bosonic systems.
 These coordinates could be supplemented with various choices of odd coordinates.
 Perhaps, the present choice  of odd coordinates looks more clear in quaternionic terms
 \beq
 \bex=\xi^1+\xi^2j=\frac{{\overline{\bet}}{\bf z}}{|z|},\qquad {\bf
 z}=z^1+z^2j.
 \eeq
Calculating the Poisson brackets among these functions and
restricting them on the level surface (\ref{const}), we shall get
the Poisson brackets on the reduced phase space
\beq\{p_{i},x_{j}\}=\delta_{ij},
\quad\{p_{i},p_{j}\}=\varepsilon_{ijk}\left(s+\frac{\imath}{2r}\xi\widehat
x\bar{\xi}\right)\; \frac{x_k}{r^3}, \quad
\{p_{i},\xi^\alpha\}=-\frac{\imath}{2r^2}\varepsilon_{ijk}x_j(\xi
\widehat\sigma_k )^\alpha , \quad \{{\xi}^\alpha
,\bar\xi^\beta\}=\delta^{{\alpha}\bar\beta}\;. \eeq
The reduced
Hamiltonian and supercharges look as follows:
 \beq {\cal
H}^{red}={r}\left[\frac{ p_i^2}{2}+\frac{s^2}{2r^2}+ s\frac{(\imath\xi\widehat
x\bar{\xi})}{2r^3} +\frac{(\imath\xi\bar{\xi})^2}{8r^2}\right],\qquad
Q_\alpha=\sqrt{r}\left[ p_r \bar{\xi}^\alpha -\imath\frac{({\widehat
{\cal J}}\bar{\xi})_{\alpha} }{r}\;\right],
\label{hrf}\eeq
while the reduced constants of motion (\ref{j}) and (\ref{r})
 take the form
\beq {\cal J}_i=
\varepsilon_{ijk}x_jp_{k}+\left(s+\frac{\imath}{2r}\xi\widehat
x\bar{\xi}\right)\frac{x_i}{r},\qquad R_+=R_1+
iR_2=-\imath\bar\xi_1\bar\xi_2,\quad R_3=-\frac{\imath \xi\bar\xi}{2}.\eeq
Here
and further below we use, for any $A_i$,  the notation ${\widehat
A}\equiv A_i\widehat\sigma_i $  and  $A_r={A_{i}x_{i}}/{r}$.\\
The Poisson brackets of these generators with the coordinates of
the reduced phase space look as follows: \beq
\{{\cal J}_{i},p_{j}\}=-\varepsilon_{ijk}p_{k}\;,\;
\{{\cal J}_{i},x_{j}\}=-\varepsilon_{ijk}x_{k}\;,\;
\{{\cal J}_{i},\xi^{\alpha}\}=\frac{i}{2}(\xi\widehat\sigma_i)^{\alpha}\;,\;
\eeq
\bea \{R_{i},x_{j}\}= \{R_{i},p_{j}\}=0,\;
\{R_{+},\xi^{\alpha}\}=-i\varepsilon_{\alpha\beta}\bar{\xi}^\beta,\;
\{R_{-},\xi^{\alpha}\}=0,\;
\{R_{3},\xi^{\alpha}\}=-\frac{\imath}{2}\xi^{\alpha}. \eea
In the given form the Hamiltonian has a canonical structure (in the sense that it is quadratic on momenta),
but the Poisson brackets are non-canonical. For a better understanding of the structure of the system
it is convenient, by a  redefinition of momenta, to transform  the Poisson bracket to the canonical
(in the absence of Dirac monopole) one
\beq
P_i=p_i+\frac{\imath}{2r^2}\varepsilon_{ijk}x_j(\xi
\widehat\sigma_k \bar\xi)\;\;:
\eeq
\beq
\{P_{i},x_{j}\}=\delta_{ij},\quad
\{P_{i},P_{j}\}=s\varepsilon_{ijk}\frac{x_k}{r^3}, \quad
\{P_{i},\xi^\alpha\}=0  ,\quad \{{\xi}^\alpha
,\bar\xi^\beta\}=\delta^{{\alpha}\bar\beta}\;. \label{canonical}\eeq
In these terms,
the $so(3)$ generators ${\cal J}_i$ take the form
\beq
{\cal J}_i\equiv J_i+\frac{\imath}{2}(\xi\widehat\sigma_i\bar{\xi}),\qquad J_i\equiv
\varepsilon_{ijk}x_jP_{k}+s\frac{x_i}{r},
\eeq
and the  Hamiltonian  looks as follows:
\beq {\cal
H}^{red}={r}\left[\frac{ P_i^2}{2}+\frac{s^2}{2r^2}+\frac{(\imath\xi{\widehat
J}\bar{\xi})}{2r^2} -\frac{(\imath\xi\bar{\xi})^2}{8r^2}\right].
\eeq
The supercharges read
\beq
Q_\alpha=\sqrt{r}\left[ P_r \bar{\xi}^\alpha -\imath\frac{({\widehat
{\cal J}}\bar{\xi})_{\alpha} }{r}\;\right]=\sqrt{r}\left[ \left(P_r+
\frac{3\imath (\imath\xi\bar\xi)}{2r}\right) \bar{\xi}^\alpha
-\imath\frac{({\widehat
J}\bar{\xi})_{\alpha} }{r}\;\right].
\label{hrf1}\eeq
Thus, we got the three-dimensional ${\cal N}=4$ supersymmetric mechanics,
specified by the presence of Dirac monopole. Its Hamiltonian, in contrast with the supercharges, looks quite simple.
But, actually, this model is quite specific, since its configuration space is non-constant, namely,
it is equipped with the metric $ds^2=(d{\bf r})^2/r$. Actually, it is not only a non-constant space,
but it has a conic singularity at the origin of the coordinates. However, on can construct, in a similar manner, the
${\cal N}=4$ supersymmetric mechanics on a three-dimensional euclidean space, as well as on the generic
three-dimensional
conformally-flat  spaces. For this purpose we should choose, as the initial system, the
four-dimensional supersymmetric mechanics on conformally-flat spaces.

\subsection*{Conformally-flat case}
Let us consider the reduction
of the ${\cal N}=4$ supersymmetric mechanics which lives on a four-dimensional space
equipped
with the conformally-flat metric
\beq
ds^2={G}(z,\bar z) dz^\alpha d{\bar z}^\alpha .
\label{cm}\eeq

The ${\cal N}=4$ supersymmetric mechanics is defined by the following
Hamiltonian   and  supercharges:
\beq
Q_1=\frac{1}{\sqrt{G}}\left(\Pi_1\eta^1+\bar{\Pi}_2\bar\eta^2\right),\quad
Q_2=\frac{1}{\sqrt{G}} \left(\Pi_2\eta^1-\bar{\Pi}_1\bar\eta^2\right),
 \label{superccf}
\eeq
\beq
 {\cal H}=\frac{\left(\Pi_{\alpha}-(\bar\eta\bar\partial\log G)\bar\eta^\alpha\right)
 \left(\bar{\Pi}_{\alpha}-(\eta\partial\log G)\eta^\alpha\right)}{2G}-
 \frac{\partial_\alpha\bar\partial_\alpha G}{4G}
(\imath\eta\bar{\eta})^2. \label{hamcf} \eeq
Here
$$
\Pi_{\alpha}\equiv\pi_{\alpha}+\imath\frac{\partial_\alpha\log G}{2}(\imath\eta\bar{\eta})
.
$$
In order to have the possibility to perform the Hamiltonian reduction by  the  generator
$J_0$ (\ref{generator}), the metric(\ref{cm})  should be invariant under the transformation (\ref{sim0}).
This means that the conformal factor  $g(z,\bar z)$ has to depend solely on
$U(1)$ invariant functions  $x_{i}$,
which are given by the expression (\ref{coord}): $G=G(x_i )$.

Repeating the Hamiltonian reduction procedure performed
in the previous Section,
 we shall get the
 three dimensional supersymmetric mechanics whose configuration space is equipped with the metric
\beq ds^2={ g}dx_idx_i,\qquad { g}\equiv \frac{G}{r}. \eeq The
connection components and scalar curvature of this metric look as
follows: \beq
\Gamma^i_{jk}=\Gamma_j\delta_{ik}+\Gamma_k\delta_{ij}-\Gamma_i\delta_{,k},
\qquad {\cal
R}=-\frac{4\partial_i\Gamma_i+2\Gamma_i\Gamma_i}{g},\qquad\quad
\Gamma_i\equiv\frac{\partial_i g}{2g}. \eeq
 The reduced supercharges are given by the following expressions:
\beq Q_\alpha=\frac{1}{\sqrt{g}}\left[
 \left(P_r +{\imath}({\Gamma_r}+\frac{2}{r})\Lambda_0-\frac{
 ({\vec r}\times{\vec\Gamma})\cdot {\vec\Lambda} }{r} \right)\bar\xi^\alpha-
\imath\frac{({\widehat
J}\bar{\xi})_{\alpha}}{r}
\;\right].
\label{3sch}\eeq
Here and in the following we use the notation and identities
\beq
\Lambda_0=(\imath\xi\bar\xi), \quad \Lambda_i=(\imath\xi\widehat\sigma_i\bar\xi )\;:
\quad \Lambda_i\Lambda_j=-\delta_{ij}\Lambda_0^2\;, \quad \Lambda_i(\widehat\sigma_j\bar\xi)_\alpha
=\left(-\delta_{ij}\Lambda_0+\imath\varepsilon_{ijk}\Lambda_k\right)\bar\xi^\alpha\;.
\eeq
The  reduced Hamiltonian looks as follows: \beq{\cal
H}=\frac{{\vec P}^2}{2g}+\frac{s^2}{2gr^2} +\frac{{\vec \Lambda}\cdot\vec
{\cal V}}{g} -(div\vec\Gamma -{\vec
\Gamma}^2+\frac{2\Gamma_r}{r}+\frac{2}{r^2})\frac{\Lambda^2_0}{2g}
\label{3sh}\eeq
where
 \beq \vec{\cal V}(\vec P, {\vec r}, s)=\frac{\vec J}{r^2}+\frac{2(\vec J \cdot\vec\Gamma)\vec r}{r}+
 {\vec \Gamma}\times{\vec P} -\frac{s}{r}\vec\Gamma
 \eeq
 For the $so(3)$ invariant
metric, $g=g(r)$, the Hamiltonian takes a quite simple form
 \beq {\cal H}=\frac{{\vec P}^2}{2g} +\frac{s^2}{2gr^2} + \left({\Gamma}+\frac{1}{r}\right)\frac{{\vec
 J}\cdot{\vec\Lambda}}{2rg}
-\left(\Gamma'-\Gamma^2+\frac{4\Gamma}{r}+\frac{2}{r^2}\right)\frac{\Lambda^2_0}{2g}
\eeq where $\Gamma\equiv \Gamma_r=d\log g(r) /(2dr)$,
$\Gamma'=d\Gamma/dr$.

The above-obtained, explicit
expressions for the supercharges (\ref{3sch}) and the Hamiltonian (\ref{3sh})
are the main results of our paper.
Since the system is  formulated in  canonical coordinates,
 it could be immediately considered
at the quantum mechanical level. For this purpose  we should replace
the  Grassman coordinates $\xi^\alpha$
by the four-dimensional Euclidean gamma-matrices $\widehat\gamma^\alpha
 = (\widehat\gamma^\alpha +\imath\widehat\gamma^{\alpha +2})/\sqrt{2}$,
 and the momenta  variables $P_i$ by the
momenta operators ${\widehat P}_i= -\imath\partial_i+ sA_i(x)$
(where $A_i$ is the potential of the Dirac monopole).

Let us draw the reader's attention to the presence of the spin-orbit coupling term
$\vec J\cdot\vec\Lambda$ and the vanishing of the explicit
 dependence of the Hamiltonian from the monopole number $s$ in the $so(3)$ symmetric case.
Even in the Euclidean space ($g=1$, $\Gamma=0$), and in the absence
of a magnetic monopole  field ($s=0$), the Hamiltonian has non-zero fermionic terms.
Hence, these terms could be interpreted as an interaction energy
of the neutral particle spin with the external field.
Notice also, that the angular part of the constructed system is the
two-dimensional mechanics obtained by the Lagrangian
reduction of the initial system in \cite{root}.

\subsection*{Comparison with previous work}
As we mentioned in the Introduction, the construction presented in this work is the Hamiltonian counterpart of the reduction
performed in \cite{root}, but with a different choice of odd coordinates. Let us  show  which Poisson brackets
arise in the original construction.
Namely, let us choose, instead of (\ref{odd}), the following
  odd coordinates:
   \beq
   \xi^{\alpha}=\frac{\bar{z}^1}{|z^1|}\eta^{\alpha}
   \label{oxi}\eeq
    as it was suggested in \cite{lectures}.

In these terms the reduced
Poisson brackets are defined by the relations
\beq\{p_{i},x_{j}\}=\delta_{ij}, \quad
\{p_{i},p_{j}\}=s\; \varepsilon_{ijk}\frac{x_k}{r^3}-\imath{R}_{ij \alpha
\bar{\beta}}\xi^\alpha \bar{\xi}^\beta ,\quad
\{p_{i},\xi^\alpha\}={\Gamma}^{\alpha}_{i\beta}\xi^\beta\;,\quad
 \{\bar{\xi}^\beta\;,\xi^\alpha\}=\delta^{\bar{\beta}\alpha}.\eeq
Here
\beq{\Gamma}^{\alpha}_{i\beta }=\frac{\imath}{2}A_{i}\delta^{\alpha}_{\beta}
\qquad {R}_{ij \alpha \bar{\beta}}=\frac{1}{2}F_{ij}\delta_{\alpha\bar{\beta}},\qquad
A_{i}=- \frac{\varepsilon_{ij 3}x_{j}}{r(r+x_3)},\eeq
i.e. $F_{ij}$ and $A_i$ are, respectively, the strength and the vector potential of the magnetic field
of the  Dirac monopole.

 In contrast with (\ref{odd}), the functions (\ref{oxi}) are singular in the line $z^1=0$, and,
in terms of the reduced space, on the semiaxis $x_3=-r$, i.e.  on the ``Dirac string".
Thus, in order to cover the whole space, we should introduce another set of
odd coordinates,
$\tilde{\xi}^{\alpha}=\frac{\bar{z}^2}{|z^2|}\eta^{\alpha}$, which
are regular on the line $z^1=0$, but singular on  $z^2=0$. These
two sets of local coordinates are related as follows:
$\tilde{\xi}^\alpha =\exp{(\imath\gamma)}\xi^\alpha $, where
$\exp{(\imath\gamma)}=\frac{z^1|z^2|}{z^2|z^1|}$,
$\gamma\in[0,4\pi)$.
Upon this choice of odd coordinates, the vector potential $A_i$
appearing in the Poisson brackets looks as follows:
$A_{i}= \frac{\varepsilon_{ij
3}x_{j}}{r(r-x_3)}$.

 We could  get the (twisted) canonical Poisson brackets (\ref{canonical}) by the following
 redefinition of momenta:
 \beq
P_i=p_i+\frac{1}{2}A_i(x)\Lambda_0.
 \eeq
 In these terms the Hamiltonian is again given  by the expression (\ref{hrf}),
 i.e. it is free from singularities on the Dirac string.
 However,  the supercharges remain singular. Moreover,  on the intersection of the
  (super)charts  the (two sets of) reduced supercharges are not equal
  to each other, but differ in the phase factor, which has no impact on the Hamiltonian.
  In other words, the supercharges are not scalar functions in this picture.

An important remark is that even for $s=0$, i.e. in the absence of Dirac monopole,
this singularity appears in the reduced Poisson bracket.
Nevertheless, this choice of coordinates is appropriate  if we reduce ourselves to
the two-dimensional system, as it was done in \cite{root}. Indeed,
this reduction assumes the choice of the bosonic coordinate
$w=z^1/z^2$; hence, the resulting system has the topology of sphere
$S^2=\DC P^1$ and, consequently, it is covered by two charts.
 Also, the  odd coordinates (\ref{oxi}) are quite
convenient, when we reduce to three dimensions the ${\cal
N}=4$ supersymmetric mechanics on K\"ahler spaces. We  are
planning to  present these systems elsewhere.


\subsection*{Conclusion}
In this paper we performed the Hamiltonian reduction of the
 simplest, $(2|2)_{\DC}$-dimensional ${\cal N}=4$
supersymmetric mechanics with flat and conformally-flat configuration spaces to
the $(3|4)_{\DR}$ dimensional ones with flat and conformally-flat phase configuration spaces.
We formulated the system in  canonical coordinates, so that it could be immediately considered
at the quantum mechanical level.
Let us mention  the appearance, in the reduced system, of the Dirac monopole magnetic field,
and of a specific spin-orbit interaction term
 mixing the momenta and
Grassmann variables. Further reduction of this system to three dimensions
should yield a system where the spin-orbit coupling term still appears,
but the Dirac monopole field is transformed
into some non-singular magnetic field (including, as a particular case,
the constant magnetic field). Hence, the constructed system could have an application
in condenced matter physics. For example, one can hope that it will be useful in the study of
the spin-Hall effect, which was observed
experimentally very recently \cite{spin-hall}. This phenomenon has been proposed to occur, as a result
of the spin-orbit coupling term  of the electron in the initial Hamiltonian.
We recall that the classical Hall effect arises
physically from a velocity dependent force, such as  the Lorentz force,
whereas another velocity dependent force in condensed matter
systems is the SO coupling force \cite{i,ii}. Thus, in
finite-size electron systems  the presence of some kind of spin-Hall effect can
be due to the interplay between the spin-orbit coupling (generating a kind
of Lorentz force)  and the  edge of the device \cite{
noish,qse,iii,iv,v}, analogously to what happens in the Hall
effect.

We have found that the constructed system
has a quite unusual structure of supercharges, and it is free of singularities
(except for the one in the coordinate origin).
It seems that the constructed system is
the generalization of the the quantum mechanics suggested
in \cite{dejohnge} to curved spaces.
We have restricted ourselves to the reduction of ${\cal N}=4$ supersymmetric mechanics, though,
in a completely similar way, one can construct the three-dimensional ${\cal N}=8$ supersymmetric mechanics,
reducing the four-dimensional system suggested in \cite{sutulin}. Also, one can apply the same technique
to  the ${\cal N}=4,8$ four-dimensional supersymmetric mechanics on K\"ahler spaces.
In particular, in this way one can construct the ${\cal N}=4$ supersymmetric
(repulsive) MIC-Kepler system, performing the Hamiltonian reduction of the ${\cal N}=4$
supersymmetric particle on the Taub-NUT space with negative mass.
The connection  between the corresponding bosonic systems has been established in
\ref{horvathy2}.

It also appears that the procedure of Hamiltonian reduction
could explain the freedom in the fermion-boson coupling observed in two-dimensional systems with
non-linear chiral multiplet \cite{ncs1}. These works are currently in progress and will be published elsewhere.

Another perspective development is in the construction of the five- dimensional supersymmetric mechanics
specified by the presence of a $SU(2)$ Yang monopole (instanton) from the
eight- dimensional supersymmetric systems (without monopoles).
For this purpose one should perform the Hamiltonian reduction by the $SU(2)$ group action,
related with the second Hopf map. The bosonic counterpart of this procedure is
widely known in classical and quantum mechanics. For example, by means of such a reduction procedure the
five-dimensional Coulomb problem  with $SU(2)$ Yang monopole has been constructed from the eight-dimensional
oscillator in \cite{iwayi}. Extending this procedure to the supersymmetric system on the eight-dimensional
conformally-flat case, one can construct the expected five-dimensional  supersymmetric system.\\

{\large Acnowledgments.} We would like to thank Sergey Krivonos  for valuable discussions and
useful remarks, and the Referee for a useful question, and for
drawing our attention to Ref.\cite{horvathy2}. This work has been supported in part by the European Community Human
Potential Program under contract MRTN-CT-2004-005104 ``Constituents,
fundamental forces and symmetries of the universe''  and by the grants
NFSAT-CRDF ARPI-3228-YE-04 and INTAS-05-7928.

\end{document}